\documentclass[twocolumn,eqsecnum,showpacs,showkeys,amsmath,amssymb,nofootinbib,superscriptaddress,floatfix]{revtex4}

\usepackage{graphicx}
\usepackage{subfigure}
\usepackage{bm}

\makeatletter
\newcommand\erfc{\mathop{\operator@font erfc}\nolimits}
\def\slashchar#1{\setbox0=\hbox{$#1$}
   \dimen0=\wd0 \setbox1=\hbox{/} \dimen1=\wd1
   \ifdim\dimen0>\dimen1 \rlap{\hbox to \dimen0{\hfil/\hfil}} #1
   \else  \rlap{\hbox to \dimen1{\hfil$#1$\hfil}} / \fi}

\makeatother

\begin{document}
\title{Viscous evolution of the rapidity distribution of matter created 
in relativistic heavy-ion collisions\footnote{Supported by 
Polish Ministry of Science and Higher Education under
grant N202~034~32/0918}}
\author{Piotr Bo\.zek}
\email{Piotr.Bozek@ifj.edu.pl}
\affiliation{Institute of Physics, Rzesz\'ow University, PL-35959 Rzesz\'ow, Poland}
\affiliation{The H. Niewodnicza\'nski Institute of Nuclear Physics,
PL-31342 Krak\'ow, Poland}
%

\begin{abstract}
Longitudinal hydrodynamic expansion of
  the fluid created in relativistic 
heavy-collisions  is considered taking into account shear viscosity.
 Both a on-vanishing viscosity and a soft equation
of state  make particle distributions in  rapidity narrower. 
The presence of viscosity  has dramatic 
consequences on the value of the initial energy density. 
The reduction of the longitudinal work and dissipative processes 
due to the  shear viscosity, 
increase the total entropy and the particle multiplicity at central 
rapidities. The total energy in the collision, dominated by the 
longitudinal motion, is conserved. Viscous corrections make 
the longitudinal velocity of the fluid  to stay close 
to the Bjorken scaling $v_z = z/t$ through the evolution. 
At the freeze-out   viscous corrections are the strongest
 for non-central rapidities.
\end{abstract}



\maketitle

\section{Introduction}
Properties of  hot and dense strongly interacting  matter can be 
studied in  ultrarelativistic nuclear collisions. 
The modeling of the evolution of the dense collective  phase is most 
commonly undertaken within the framework of  relativistic
 fluid dynamics \cite{Huovinen:2003fa,Kolb:2003dz,Nonaka:2007nn}. 
A clear evidence of a collective behavior 
of the system created in a collision is given by the observation 
of  transverse elliptic flow  of the produced particles. Collective 
flow arises naturally during a hydrodynamic evolution. In
 ultrarelativistic heavy-ion collisions the movement of the 
matter at the initial stage is dominated by the expansion in the 
longitudinal direction. 
 Most 
of the experimental data at the Relativistic Heavy- Ion
 Collider (RHIC)
 are restricted to the central 
rapidity region. Therefore, hydrodynamic models often assume a 
simplified geometry of the fireball with a Bjorken boost invariant 
flow in the longitudinal direction and concentrate on the dynamics in the
 transverse directions
with  azimuthal symmetry for central collisions
 \cite{Baym:1983,Rischke:1996em,Kataja:1986}
 or azimuthally asymmetric geometry for
 collisions of nuclei at non-zero impact parameters 
\cite{Ollitrault:1992,Kolb:2000sd,Huovinen:2001cy,Teaney:2000cw}. Only 
few calculations consider a fully three-dimensional evolution of 
the fluid \cite{Nonaka:2000ek,Hirano:2001yi,Hirano:2002ds}. 
 Results of the hydrodynamic evolution
 are sensitive to the chosen equation of state. Values of the 
Hanbury Brown-Twist
 (HBT) radii observed at RHIC \cite{Adams:2004yc} 
seem to exclude equations of state
 with a strong first order phase transition, or even with  a more prominent
 soft point. The hydrodynamic 
evolution depends on  the initial temperature 
and its profile, on the chosen equation of state, 
and on the freeze-out temperature. 
Particles emitted
 at the hydrodynamic freeze-out can still rescatter, which modifies somewhat
 their spectra and elliptic flow
 \cite{Bass:2000ib,Teaney:2000cw,Hama:2005dz,Nonaka:2006yn}. 

Instead of a
 fully three-dimensional calculation only longitudinal expansion
 of the matter created in a collision 
can be considered 
\cite{Baym:1983,Kajantie:1983}. 
Recently the rapidity distributions of pions and 
kaons produced in $\sqrt{s} = 200$AGeV Au-Au collisions measured by the 
BRAHMS Collaboration \cite{Bearden:2004yx}
 have been analyzed in a 1+1 dimensional
 hydrodynamic model \cite{Satarov:2006iw}. 
The final rapidity distribution of
 mesons has been found to be sensitive to the initial energy density
distribution in rapidity and to the chosen equation of state.  
Experimental data indicate that the boost invariant Bjorken scaling 
solution is not realized at RHIC energies. Results of the 1+1 
dimensional longitudinal fluid dynamics for the meson rapidity 
distributions favor a soft equation of state and require a Gaussian 
initial energy density distribution in the space-time
 rapidity \cite{Satarov:2006iw}.

Analysis of the elliptic flow in the momentum distribution
 of  particles as a function of the initial eccentricity 
of the source points to the possibility of a noticeable effect of 
the shear viscosity of the fluid  \cite{Aguiar:2001ac,Drescher:2006ca}.
 Quantitative studies have to 
take into account many effects, initial conditions, the equation of state,
 the freeze-out temperature, possible final rescattering,
 therefore estimates of
 the viscosity coefficient are still under debate 
\cite{Song:2007fn,Lacey:2007na}. 
Theoretical estimates of the ratio of the shear viscosity coefficient
 $\eta$ to the entropy density
$s$ range from a conjectured lower bound $\eta/s=1/4\pi$ \cite{Kovtun:2004de} 
to $\eta\simeq s$
 \cite{Venugopalan:1992hy,Arnold:2000dr}.

The role of the shear viscosity in the dynamics is most
important during the early evolution of the system, when 
the velocity gradients are the largest. Gradients of the Bjorken
 flow give rise to corrections of the pressure tensor in the fluid.
 The transverse pressure increases, and the fluid expands faster in 
the transverse directions, this leads to stronger transverse flow and 
also to the  saturation of the elliptic flow 
\cite{Song:2007fn,Teaney:2003kp,Baier:2006sr,Baier:2006gy,Chaudhuri:2006jd,Romatschke:2007jx,Hirano:2005xf}. All recent calculations
 using viscous relativistic hydrodynamics assume boost invariant Bjorken
 flow in the longitudinal direction and study the transverse 
development of the fluid in azimuthally symmetric or asymmetric 
conditions. Longitudinal pressure is reduced, and hence so is the 
longitudinal flow of the fluid. The fluid cools slower, at least  until 
substantial transverse flow builds up. Weaker longitudinal expansion
 and entropy production due to dissipative evolution require an 
adjustment of the initial entropy (temperature). As a result, the 
lifetimes of the plasma in the viscous and ideal fluid evolutions are 
similar. Finally, let us note that viscous corrections to the distribution
 functions at the freeze-out modify the spectra, and the HBT radii
 \cite{Romatschke:2007jx,Muronga:2004sf}. 

Enhanced transverse pressure and the resulting modification 
of the spectra at central rapidity, raise the question of possible 
modifications of the fluid dynamics in the longitudinal direction
 due to viscosity. Since the boost invariant scaling solution
 is not applicable at RHIC energies, quantitative description of the
 energy flow and entropy production in the fireball should take into
 account a fully three dimensional geometry. Such a task within the 
dissipative fluid dynamics has not been accomplished yet. In the following
 we consider a simpler problem of the evolution of a longitudinally 
expanding non-boost invariant fluid with viscosity. One expects that
 reduced longitudinal work in a viscous fluid generates narrower particle 
distributions in rapidity. The author is aware of only one, two-decades 
old work, considering this problem \cite{chu:1986}, where first-order viscous
 hydrodynamics has been applied.
 At lower energies no strong effect
 of viscosity on the longitudinal expansion has been observed \cite{chu:1986}.

 In
 the following we use BRAHMS data \cite{Bearden:2004yx}
 to constraint meson rapidity
 distributions after freeze-out. Solving viscous hydrodynamics in the 1+1
 longitudinal geometry, we find a significant reduction of the initial 
energy density when viscosity is taken into account. Also the longitudinal
 flow is modified, viscosity reduces the flow and counteracts the 
acceleration due rapidity gradients of the  pressure.
At the Large Hadron Collider (LHC) energies
 we find that the modification of the 
longitudinal dynamics
due to shear viscosity leads to an increase of the rapidity range where
 the Bjorken scaling flow applies.

\section{Longitudinal hydrodynamic equations with shear viscosity}

We consider a baryon-free fluid with non-zero shear viscosity. 
The energy-momentum tensor is the sum of the ideal fluid component 
and the  shear tensor $\pi^{\mu \nu}$
\begin{equation}
T^{\mu\nu}=(\epsilon+p)u^\mu u^\nu -g^{\mu \nu}+\pi^{\mu\nu} \ ,
\label{eq:emtensor}
\end{equation}
$\epsilon$ and $p$ are the local energy density and pressure of the fluid, and
\begin{eqnarray}
u^\mu &= & \gamma(1,0,0,v) \nonumber \\
& =& (\cosh Y ,0,0, \sinh Y) 
\end{eqnarray}
is the four-velocity of the fluid element ($\gamma=1/\sqrt{1-v^2}$),
 and $Y=\frac{1}{2}\ln \left(\frac{1+v}{1-v}\right)$ is its rapidity.
The energy density and the pressure are related by the 
equation of state.
The energy density $\epsilon(t,z)$
 and the longitudinal velocity component $v(t, z)$ 
are functions of time $t$ and the beam
axis coordinate $z$. Instead of the time  and the $z$ coordinate it  is
preferable to use the proper time $\tau=\sqrt{t^2-z^2}$
 and the space-time rapidity
\begin{equation}
\theta=\frac{1}{2}\ln\left( \frac{t+z}{t-z}\right) \ .
\end{equation}
Hydrodynamic equations $\partial_\mu T^{\mu \nu} =0 $ can be written as
\cite{Muronga:2006zw,Baier:2006um}
\begin{equation}(\epsilon+p) Du^\mu= 
\nabla^\mu p -\Delta^\mu_\nu \nabla_\alpha 
\pi^{\nu\alpha}+\pi^{\mu\nu}Du_\nu 
\label{eq:vh1}
\end{equation}
and
\begin{equation}
D\epsilon= -(\epsilon+p) \nabla_\mu u^\mu +\frac{1}{2} \pi^{\mu\nu} 
 < \nabla_\mu u_\nu > \ ,
\label{eq:vh2}
\end{equation}
where
\begin{equation}
D=u^\mu\partial_\mu=\cosh(Y-\theta)\partial_\tau
+\frac{\sinh(Y-\theta)}{\tau}\partial_\theta \ ,
\end{equation}
\begin{equation}
< \nabla_\mu u_\nu >=
\nabla_\mu u_\nu+\nabla_\nu u_\mu-
\frac{2}{3}\Delta_{\mu\nu}\nabla_\alpha u^\alpha \ ,
\end{equation}
\begin{eqnarray}
\nabla^\mu&=&\Delta^{\mu\nu}\partial_\nu\nonumber \\
& =& (-\sinh Y {\cal K},-\partial_x,-\partial_y,-\cosh Y{\cal K})
\end{eqnarray}
with 
$\Delta_{\mu\nu}=g_{\mu\nu}-u_\mu u_\nu$ and
\begin{equation}
{\cal K}=\sinh(Y-\theta)\partial_\tau+\frac{\cosh(Y-\theta)}
{\tau}\partial_\theta \ .
\end{equation}
The equations of the  second order viscous hydrodynamics
  (\ref{eq:vh1}) and (\ref{eq:vh2}) are supplemented with a
dynamic equation for the stress tensor 
\cite{Baier:2006sr,Muronga:2004sf,Muronga:2001zk,IS,Baier:2006um,Muronga:2003ta}
\begin{equation}
\tau_\pi \Delta^\mu_\alpha\Delta^\nu_\beta \pi^{\alpha\beta}+\pi^{\mu\nu}=
\eta < \nabla^\mu u^\nu >-2\tau_\pi \pi^{\alpha(\mu}\omega^{\nu)}_\alpha
\end{equation}
$\eta$ is the shear viscosity coefficient and $\tau_\pi$ is 
the relaxation time of the stress tensor.
 $\omega^{\mu\nu}=\Delta^{\mu\alpha}\Delta^{\nu\beta}
\left(\partial_\alpha u_\beta-\partial_\beta u_\alpha\right)$ 
is the vorticity of the fluid; it is zero for the longitudinal 
flow considered here. The relaxation time and the 
viscosity coefficient can be estimated from microscopic models,
 considering equilibration processes 
\cite{Kovtun:2004de,Venugopalan:1992hy,Arnold:2000dr,Muronga:2007qf}. 
In this work we 
take several values for the ratio of the viscosity coefficient 
to the entropy $\eta/s$, and we drop 
 viscosity effects for temperatures
below $130$MeV, the relaxation time is \cite{Venugopalan:1992hy}
$\tau_\pi/\eta = 6/T s $, unless specified otherwise 
($T$ is the local temperature).
 For short  relaxation times  the stress tensor 
relaxes fast and stays close to the Navier-Stokes
 value $\pi^{\mu\nu}=\eta<\nabla^\mu u^\nu>$. 
The dependence on the initial value of the stress tensor and 
on the relaxation time is discussed in Section \ref{sec:stressini}.

For a fluid expanding only in the longitudinal direction, 
with the energy density and the velocity constant in the transverse plane, 
the stress tensor can be written using one scalar function $\Pi$
\begin{equation}
\pi^{\mu\nu}=\left(\begin{array}{cccc} -\sinh^2 Y & 0 & 0 & -\sinh Y \cosh Y\\
0 & \frac{1}{2} & 0 & 0 \\
0 & 0 & \frac{1}{2} & 0 \\
-\sinh Y \cosh Y & 0 & 0 & -\cosh^2 Y
\end{array}\right)\Pi
\end{equation}
and viscous  hydrodynamic equations take the form
\begin{eqnarray}
(\epsilon+p)DY&=& -{\cal K}p+\Pi DY + {\cal K} \Pi \nonumber \\
D\epsilon&=& (\epsilon +p) {\cal K} Y -\Pi {\cal K} Y\nonumber \\
D\Pi&=& (\frac{4}{3}\eta {\cal K}Y-\Pi)/\tau_\pi  \ .
\label{eq:eqsolv}
\end{eqnarray}
The above equations are solved numerically in the
 $\tau$-$\theta$  plane, starting from
 some energy density $\epsilon(\tau_0,\theta)$ 
at the initial proper time of the evolution  $\tau_0=1.0$fm/c. 
For the initial fluid rapidity we always take the Bjorken flow
\begin{equation}
Y(\tau_0,\theta)=\theta \ .
\label{eq:flowini}
\end{equation}
Assuming boost invariance, i.e.
\begin{equation}
Y(\tau,\theta)=\theta \ , \ \ \epsilon(\tau,\theta)=\epsilon(\tau) \ , \ \
p(\tau,\theta)=p(\tau)
\end{equation}
hydrodynamic equations simplify to \cite{Muronga:2001zk,Baier:2006um}
\begin{eqnarray}
\frac{d\epsilon}{d\tau}&=& -\frac{\epsilon+p-\Pi}{\tau} \nonumber \\
\frac{d\Pi}{d\tau}&=& \frac{(4\eta)/(3\tau)-\Pi}{\tau_\pi} \ .
\label{eq:bjorkenvis}
\end{eqnarray}

\section{Equation of state}
\label{sec:eos}

\begin{figure}
\includegraphics[width=.5\textwidth]{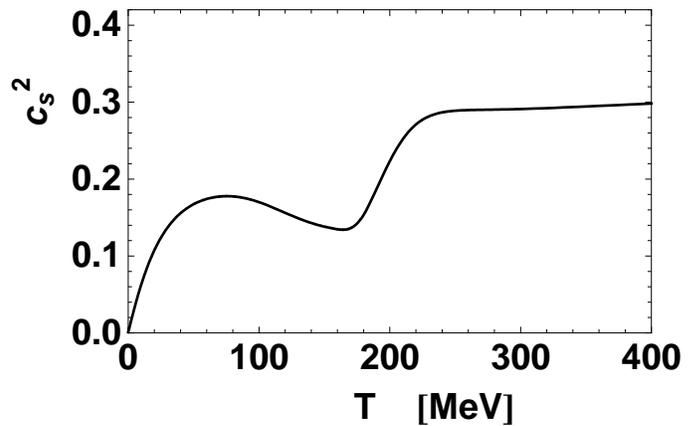}
\caption{Square of the velocity of sound as a function of the 
temperature for an equation of state interpolating between the
 hadron gas and the quark-gluon plasma expressions \cite{Chojnacki:2007jc}.}
\label{fig:cs}
\end{figure}

 The equation of state determines the evolution of the fireball. 
In the following we take a parameterization of the equation 
of state proposed by Chojnacki and Florkowski \cite{Chojnacki:2007jc}. 
It is an interpolation of the lattice data at temperatures above 
$T_c = 170$MeV and of an equation of state of noninteracting hadrons
 at lower temperatures. The two limiting formulas are joined 
smoothly with only a slight softening of the equation of state around
 the critical temperature (Fig. \ref{fig:cs}). 
This minimally softened, realistic equation 
of state has shown itself suitable for the description of the transverse 
expansion of the fluid and of the build up of the elliptic flow and gives  
reasonable HBT radii
 \cite{Chojnacki:2007jc,Chojnacki:2007rq}.  

In this section we consider the hydrodynamic 
longitudinal expansion of an ideal fluid. Equations for the 
fluid rapidity and energy density are obtained from Eqs. (\ref{eq:eqsolv})
 setting 
$\Pi  = 0$. 
This problem has been discussed recently in the 
context of heavy-ion collisions at RHIC \cite{Satarov:2006iw}.
 For completeness
 we study the effect of the equation of state we use on the
 longitudinal expansion. For that purpose we compare the evolution
 using the equation of state of Chojnacki and Florkowski with an evolution
based on  a 
relativistic gas equation of state $p=\frac{1}{3}\epsilon$.

At the initial time
$\tau_0= 1.0$fm/c the energy density is
\begin{equation}
\epsilon(\tau_0,\theta)=\epsilon_0 \exp\left(-\theta^2/(2\sigma^2)\right) \ . 
\label{eq:eini}
\end{equation} 
The initial energy density $\epsilon_0$
 and the width of the 
initial rapidity distribution $\sigma$ are parameters adjusted to 
reproduce final meson rapidity distributions. 
The freeze-out takes place at the temperature $T_f = 165$MeV, this high 
freeze-out temperature is the same as the chemical freeze-out temperature
 \cite{BraunMunzinger:2001ip,Florkowski:2001fp}. The spectrum of 
particles emitted with four-momentum $q^\mu=(E,q)$ 
 is given by the Cooper-Frye formula \cite{Cooper:1974mv,Satarov:2006iw}
\begin{equation}
E\frac{d^3N}{d^3q}=\frac{d^3N}{dy d^2q_\perp}=\frac{1}{(2\pi)^3}\int
d\Sigma_\mu q^\mu f(q^\mu u_\mu) \ ,
\label{eq:cff}
\end{equation}
$f(E)=e^{-E/T_f}$ is the thermal distribution (Boltzmann 
distribution for simplicity), and $y$ denotes the rapidity of the 
emitted particle.
 The element of the hypersurface of 
constant freeze-out temperature is 
$d\Sigma^\mu = S\ (\tau^{'}(\theta)\sinh\theta +\tau(\theta)\cosh\theta,
0,0,\tau^{'}(\theta)\cosh\theta -\tau(\theta)\sinh\theta)$, $S = \pi R_{Au}^2$ 
is the transverse area of the fireball in central collisions
and $\tau(\theta)$ is the line of constant temperature
 $T_f$ in the $\tau$-$\theta$ plane. Particle distributions
 in rapidity are obtained by integration over the  transverse momenta
$q_\perp$ in Eq. (\ref{eq:cff})
\begin{eqnarray}
\frac{dN}{dy}&=& \frac{S}{4\pi^2}\int_{-\theta_{max}}^{\theta_{max}} 
\left( \tau(\theta) \cosh(y-\theta)-\tau^{'}(\theta)\sinh(y-\theta)\right)
\nonumber \\ & & 
(2 m \xi +2 \xi^2 +m^2)\xi \nonumber \\ & & 
\exp\left( -\frac{m\cosh(y-Y_f(\theta))}{T_f}  \right) d\theta \ , 
\label{eq:dndyfree}
\end{eqnarray}
$m$ is the meson mass, $Y_f(\theta)=Y(\tau(\theta),\theta)$ is 
the fluid rapidity at the freeze-out hypersurface, and
\begin{equation}
\xi=\frac{T_f}{\cosh(y-Y_f(\theta))} \ .
\end{equation}

The above expression neglects the transverse expansion of the fluid 
at the freeze-out. This has only a small effect on rapidity distributions,
 the distribution should be narrower. Pions 
and kaons come to a large extend from secondary decays of resonances. The 
emission takes place in two stages, first an emission of a heavy
 resonance according to Eq. (\ref{eq:dndyfree}) and then the decay 
of the resonance into pions 
(kaons). The emission of resonances and their decay is also 
influenced by their transverse expansion. A hint on the 
spread in rapidity of the decay products of resonances is given
 by charge balance correlations \cite{Adams:2003kg,Bozek:2003qi,Cheng:2004zy}.
 Narrow charge balance 
functions indicate that  decay products of a resonance are only $0.5$ 
unit of rapidity away from the parent resonance, convoluting this 
distribution with the spread in rapidity of the emitted  resonances one obtains
 a distribution of half-width similar as for the emission of direct
 pions in Eq. (\ref{eq:dndyfree}).
 Since 75\% of pions come from resonances at
$T_f = 165$MeV \cite{Torrieri:2004zz},
 we  multiply the distribution from (\ref{eq:dndyfree}) 
by a factor 4 to account for all pions, direct
and from resonance decays (the factor is 1.7 for kaons).
\begin{figure}
\includegraphics[width=.5\textwidth]{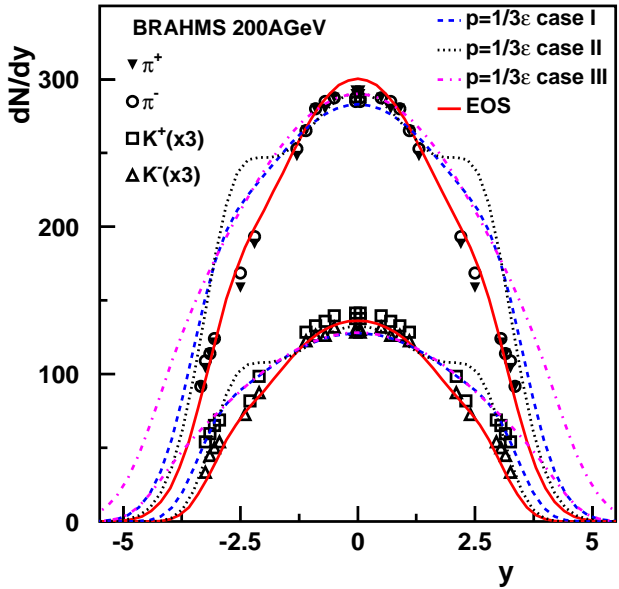}
\caption{(color online) Rapidity distribution 
of mesons calculated using a realistic equation of 
state (solid line) and using a relativistic gas
 equation of state for three different  initial
 conditions (dashed, dotted and dashed-dotted lines), Table I. 
Data are from the BRAHMS Collaboration \cite{Bearden:2004yx}.}
\label{fig:dndyfree}
\end{figure}

\begin{table}
\begin{tabular}{|l|c|c|c|} \hline 
& $\epsilon_0$ & $\sigma$& $\tau(0)$ \\
& GeV/fm$^3$ & & fm/c \\  \hline 
EOS \cite{Chojnacki:2007jc}& 16.9 & 1.05 & 14.8 \\
$p=\frac{1}{3}\epsilon$  case I & 71.5 & 1.05 & 14.8 \\
$p=\frac{1}{3}\epsilon$  case II & 102 & 0.8 & 15.7 \\
$p=\frac{1}{3}\epsilon$  case III & 50.8  & 1.5 & 13.7 
 \\
\hline
\end{tabular}
\caption{Parameters of the initial energy density distribution (\ref{eq:eini}),
for ideal fluid calculations, one 
using a realistic equation of state 
and three calculations using a relativistic gas equation of state. 
In the last column is shown the lifetime of the system until freeze-out.}
\label{table1}
\end{table}
\begin{figure}
\includegraphics[width=.5\textwidth]{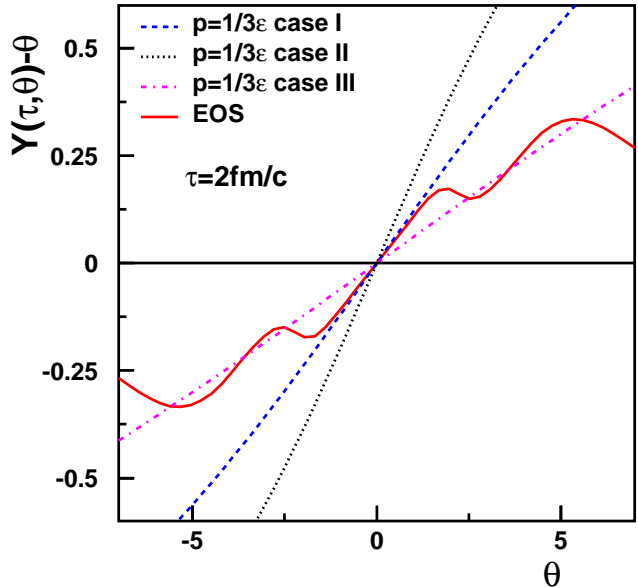}
\caption{(color online) Difference between the  flow rapidity
 of the fluid and the Bjorken value, calculated using a realistic 
equation of state (solid line) and using a relativistic gas
 equation of state for three different  initial conditions 
(dashed, dotted and dashed-dotted lines), Table \ref{table1}.}
\label{fig:flowfree}
\end{figure}

The parameters $\epsilon_0$  and $\sigma$
 have been adjusted for the calculation using a realistic
 sound velocity (Fig. \ref{fig:cs}) to reproduce the 
width and normalization of the observed pion distribution. 
The resulting meson distributions are similar as the ones  
observed experimentally. 
When using a relativistic gas equation 
of state ($p=\frac{1}{3}\epsilon$) 
one always gets a meson distribution in rapidity that
 is too wide. We present three calculations with different 
initial widths $\sigma$ and with the initial energy densities
 $\epsilon_0$ adjusted to reproduce
$dN/dy$ for central rapidities only (Fig. \ref{fig:dndyfree}) . 
The final meson distribution is much wider
 than the initial energy density distribution in all cases. 
It is due to the
breaking of the Bjorken scaling of the longitudinal
 flow, $Y(\tau,\theta)>\theta$  (Fig. \ref{fig:flowfree}). 
Fast moving fluid elements emit
mesons in the far forward and backward rapidities. Concluding this section,
 we confirm the findings of Ref. \cite{Satarov:2006iw}. 
A hard equation of state never works; a narrow 
initial distribution of the energy density in rapidity leads to a strong 
acceleration of the longitudinal flow, wider initial distributions 
 are incompatible with  the narrow final meson distributions. Only 
by imposing a softened equation of state,
 experimental pion and kaon distributions
 can be approximately reproduced. In the following we take the realistic 
equation 
of state from Ref. \cite{Chojnacki:2007jc} and study 
the effect of non-negligible shear viscosity.

\section{Dissipative longitudinal expansion}

\begin{figure}
\includegraphics[width=.5\textwidth]{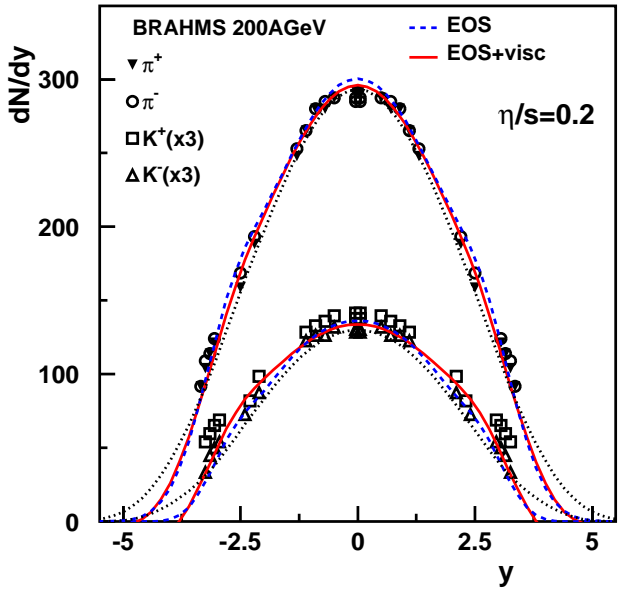}
\caption{(color online) Rapidity distribution of mesons calculated 
using a realistic equation of state and viscosity $\eta/s = 0.2$
 (solid line) and using ideal fluid hydrodynamics (dashed line).
 The dotted line denotes the results of a viscous hydrodynamic evolution,
 but neglecting the viscous corrections to the particle emission at 
freeze-out (Eq. \ref{eq:dndydissi}). Data are from the BRAHMS Collaboration 
\cite{Bearden:2004yx}.}
\label{fig:dndy}
\end{figure}

Following the hydrodynamic evolution with viscosity requires the solution
 of the  coupled equations (\ref{eq:eqsolv}), for the fluid rapidity $Y$, 
the viscous correction $\Pi$,  and the energy density $\epsilon$. 
The initial conditions are given in Eqs. 
(\ref{eq:eini}) and (\ref{eq:flowini}). For 
the initial 
viscous corrections we take $\Pi (\tau_0,\theta)=p(\tau_0,\theta)$.
 This means that at the initial time the effective 
pressure is maximally anisotropic
in the fluid rest frame, with zero longitudinal pressure. 
We vary the 
value of the shear viscosity coefficient $\eta/s$,
and in each case we adjust the parameters of the initial energy
 density distribution (\ref{eq:eini}) to reproduce BRAHMS data for the pion 
distribution in rapidity. Viscous corrections modify the momentum 
distribution functions in the fluid. We assume the same modification 
for all the particle species in the fluid
\begin{equation}
f(q)+\delta f(q)=f(q)\left(1+\frac{q_\mu q_\nu \pi^{\mu\nu}}{2T^2 (\epsilon+p)}
\right) \ ,
\end{equation}
\begin{figure}
\includegraphics[width=.5\textwidth]{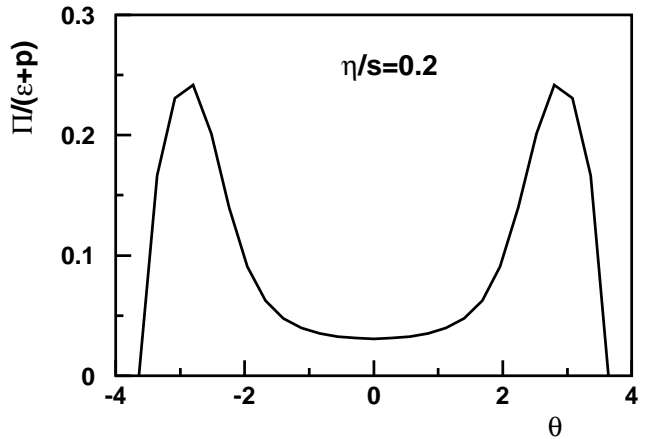}
\caption{Relative viscous corrections $\Pi/(\epsilon+p)$
at the freeze-out as a function of space-time rapidity (Eq. \ref{eq:visc}).}
\label{fig:visc}
\end{figure}
The correction to the particle distribution function is proportional
 to the ratio of the viscous correction $\Pi$  to the enthalpy
$\epsilon+p$. Viscous corrections to the 
distribution functions at the freeze-out modify the Cooper-Frye formula
\begin{equation}
\frac{dN_{visc}}{dy}=\frac{dN}{dy}+\frac{d\delta N}{dy} \ \ .
\label{eq:dndy}
\end{equation}
To the expression (\ref{eq:dndyfree}) one has to add
\begin{eqnarray}
\frac{d\delta N}{dy}& &= \frac{S}{4\pi^2}\int_{-\theta_{max}}^{\theta_{max}} 
\left( \tau(\theta) \cosh(y-\theta)-\tau^{'}(\theta)\sinh(y-\theta)\right)
\nonumber \\ & & 
\left[12 \xi^5 +5\xi^3m^2+12 \xi^4 m + \xi^2 m^3 -\sinh(y-Y_f(\theta)) 
\right. \nonumber \\ & &  \left.
(24 \xi^5 +12 \xi^3 m^2 +24\xi^4 m +4 \xi^2 m^3 +\xi m^4) \right]
\nonumber \\ & &
\frac{\Pi}{2T^2 (\epsilon+p)}
\exp\left( -\frac{m\cosh(y-Y_f(\theta))}{T_f}  \right) d\theta \ , 
\label{eq:dndydissi}
\end{eqnarray}
In Fig. \ref{fig:dndy}
 is shown the result of this procedure for the case
$\eta/s = 0.2$. Again the direct meson spectra are multiplied by a factor $4$
 for pions and $1.7$ for kaons, to account for the expected ratio of 
 all  mesons to
directly produced mesons  \cite{Torrieri:2004zz},
 at the chosen freeze-out temperature. Pion emission at 
the end of the viscous hydrodynamic evolution (solid line) is similar 
as observed experimentally. In Fig. \ref{fig:dndy}
 is also shown the meson distribution obtained using the equilibrium 
distribution at freeze-out (Eq. \ref{eq:dndyfree}) (dotted line). 
Deviations from the full result (Eq. \ref{eq:dndy}) is only noticeable 
at rapidities $3$ unit away from central rapidity. 
It can be understood as due to an  earlier freeze-out at large 
rapidities, which makes the relative viscous corrections
\begin{equation}
\frac{\Pi(\tau(\theta),\theta)}{\epsilon(\tau(\theta),\theta)
+p(\tau(\theta),\theta)}
\label{eq:visc}
\end{equation}
at freeze-out more important  (Fig. \ref{fig:visc}). 
Accordingly one expects the largest correction from viscosity 
to particle spectra, elliptic flow, and HBT radii at large rapidities.

\begin{figure}
\includegraphics[width=.5\textwidth]{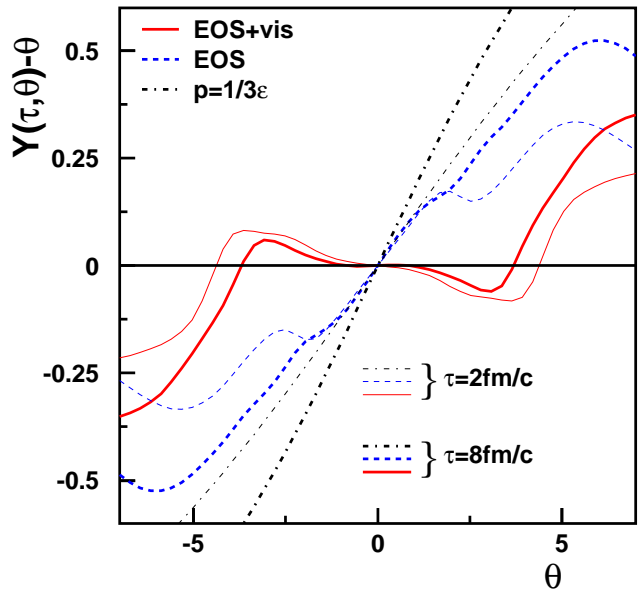}
\caption{(color online)Difference between the  flow rapidity 
of the fluid and the Bjorken value, calculated for an evolution with 
shear viscosity coefficient $\eta/s = 0.2$ (solid lines), for an ideal
 fluid with a realistic equation of state (dashed lines) and using a
 relativistic gas equation of state (case I) (dashed-dotted lines).}
\label{fig:flow}
\end{figure}
\begin{table}
\begin{tabular}{|l|c|c|c|} \hline 
$\eta/s$ & $\epsilon_0$ & $\sigma$& $\tau(0)$ \\
& GeV/fm$^3$ & & fm/c \\  \hline 
0 & 16.9 & 1.05 & 14.8 \\
0.1 & 9.8 & 1.18 & 14.1 \\
0.2 & 5.6 & 1.8 & 13.1 \\
0.3 & 4.0 & 1.86 & 12.4 \\
\hline
\end{tabular}
\caption{Parameters of the initial energy density distribution (\ref{eq:eini}) 
for hydrodynamic calculations with several value of the shear viscosity
 coefficient. In the last column is show the lifetime of the system 
until freeze-out.}
\label{table2}
\end{table}
\begin{figure}
\includegraphics[width=.5\textwidth]{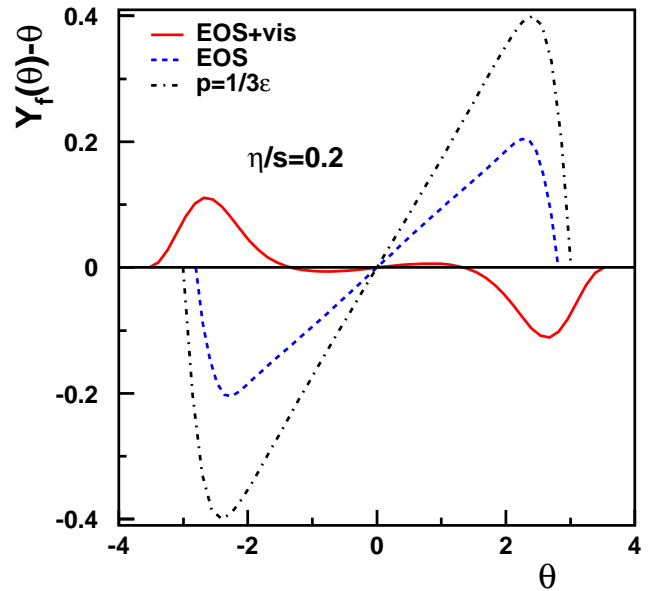}
\caption{(color online)Difference between the  flow rapidity 
of the fluid and the Bjorken value at the freeze-out hypersurface, 
calculated for an evolution with 
shear viscosity coefficient $\eta/s = 0.2$ (solid line), for an ideal
 fluid with a realistic equation of state (dashed line) and using a
 relativistic gas equation of state (case I) (dashed-dotted line).}
\label{fig:flowfreeze}
\end{figure}
\begin{figure}
\includegraphics[width=.5\textwidth]{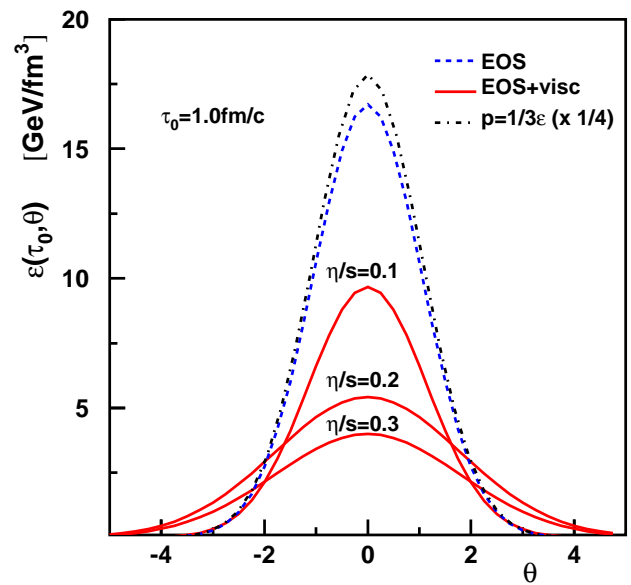}
\caption{(color online) Initial energy density distribution for 
the ideal fluid hydrodynamic evolution with a realistic equation of
 state (dashed line), for  viscous hydrodynamic evolutions 
(solid lines), and for a relativistic gas equation of state (case I) 
(dashed-dotted line).}
\label{fig:eini}
\end{figure}
We have noticed (Section \ref{sec:eos}) that longitudinal gradients of the 
pressure 
 cause the 
acceleration of the motion of the fluid in the beam direction. 
On the other hand, 
viscosity reduces the longitudinal motion of the fluid, and as a 
consequence reduces its expansion in the longitudinal direction. 
Shear viscosity prevents large gradients of the velocity field to develop. 
One can compare the rapidity of the fluid with and without shear 
viscosity (Fig. \ref{fig:flow}). 
For $\eta/s = 0.2$,
 the flow stays close to the Bjorken one during a substantial part 
of the evolution. 
At the freeze-out hypersurface the flow is still 
Bjorken-like for $|\theta|<1.8$
for $\eta/s=0.2$ (Fig. \ref{fig:flowfreeze}). The acceleration from 
pressure gradients and the deceleration from viscosity approximately
 cancel for this choice of parameters.
Reduced longitudinal expansion with viscosity requires smaller
 initial energy densities to reproduce the final meson distributions.
 In Table \ref{table2}
 are listed the initial energy densities and widths of rapidity 
distributions adjusted to reproduce the observed meson distributions
 for several values of the shear viscosity coefficient. The shape of 
the initial energy density is extremely sensitive to the dynamics of the 
longitudinal expansion (Fig. \ref{fig:eini}). Obviously the value 
of the initial 
energy density, or in other words the cooling rate from the longitudinal 
motion, is an important ingredient in the modeling of the transverse 
expansion of the fluid. Let us also note, that 
when reduced initial energy densities are imposed, 
the lifetime of the system until freeze-out is only weakly dependent 
on the viscosity coefficient.

\section{Cooling and entropy production}

\begin{figure}
\includegraphics[width=.5\textwidth]{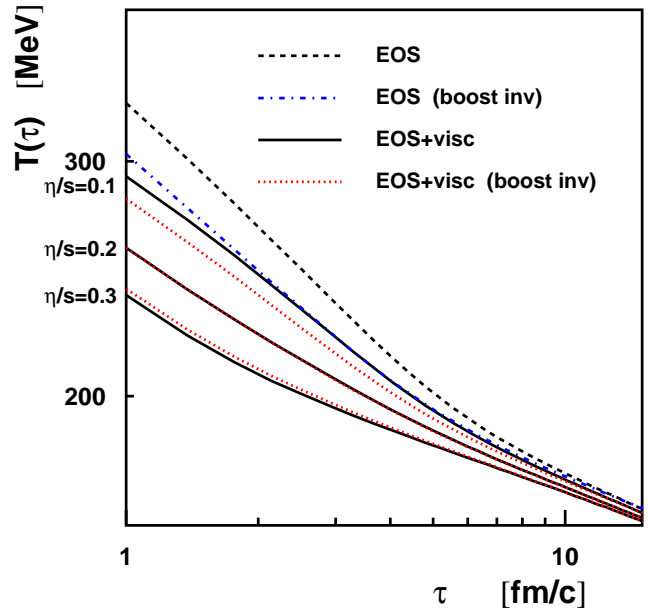}
\caption{(color online) Time dependence of the temperature at 
the center for a longitudinally expanding  ideal fluid fireball
 (dashed line) compared to 
the boost-invariant solution (dashed-dotted line). Same for the 
evolution with viscosities $\eta/s = 0.1,\ 0.2,\ 0.3$ (solid lines,
 increasing $\eta$ from top to bottom), 
and for the 
boost-invariant case with viscosity (dotted lines). For $\eta/s=0.2$ the
 solid and doted curves lie on top of each other.}
\label{fig:cool}
\end{figure}

It is instructive to compare the cooling rate in our 
solution and in the boost invariant scaling solution. Finite
 extension in space-time rapidity makes the cooling rate faster.
 On the other hand, reduced velocity of sound and shear viscosity
 reduce the longitudinal work of the pressure, and slow down the cooling. 
In Fig. \ref{fig:cool} is compared the cooling  of the central 
region of a finite system to the 
cooling in the boost invariant case (Eq. \ref{eq:bjorkenvis}). 
To compare the solution in the 1+1 dimensional system to the boost-invariant 
one, the initial temperatures are fixed so 
as to give the same lifetimes of the two systems until freeze-out.
 Boost invariant solutions underestimate the values of the initial
 temperature (energy density) and of the cooling rate 
for an ideal fluid evolution.
In a finite system additional cooling appears and the longitudinal flow 
is stronger than the Bjorken scaling flow. As the shear viscosity coefficient 
increases the velocity gradients in the dynamics are more and more constrained.
 The flow accelerates less. At $\eta/s=0.2$ the effects of the viscosity 
and space-time rapidity gradients of the pressure 
counterbalance each other and the flow is
 approximately Bjorken-like. With stronger viscosity $\eta/s=0.3$, 
the Bjorken flow 
is decelerated and the cooling is slower than for the boost-invariant solution.
It means that the gradients of the viscous correction $\Pi$ are larger 
than the gradients of the pressures itself and  applicability of second order
 viscous hydrodynamics is questionable.

Dissipative hydrodynamics conserves the total
 energy but produces entropy. The expression for the total energy of 
the system at proper time $\tau$ is
\begin{eqnarray}
E(\tau)&=& \tau S \int_{-\infty}^\infty d\theta \left[\epsilon 
 \cosh Y \cosh(Y-\theta)
\right. \nonumber \\
&+& \left. (p-\Pi)\sinh Y \sinh(Y-\theta)\right] \ .
\label{eq:etot}
\end{eqnarray}
 It is 
mainly composed of the kinetic energy of the longitudinal motion
 of the fluid. Small changes of the energy density at large rapidities
 can cause significant changes of the energy density for central rapidities. 
Softening of the equation of state and non-zero 
shear viscosity modify the dynamics 
at large rapidities, which leads to less cooling at $\theta= 0$, 
while the global
 energy of the fireball is unchanged. Dissipative processes
 driving the system locally towards equilibrium produce entropy 
\cite{IS,Muronga:2003ta,Muronga:2007qf,Elze:2001ss,Dumitru:2007qr}. 
The total entropy of the system
\begin{figure}
\includegraphics[width=.5\textwidth]{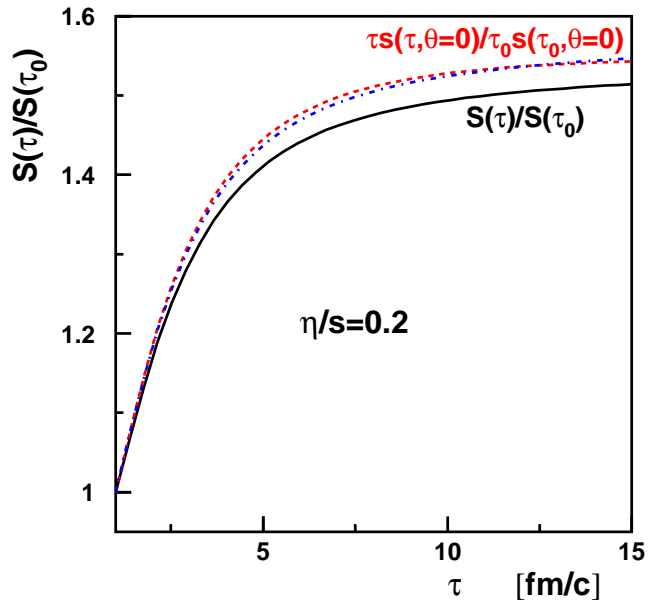}
\caption{(color online) Relative entropy production in the 
viscous hydrodynamic evolution (Eq. \ref{eq:stot}) (solid line), 
of the entropy density at central space-time rapidity (dashed line),
 and of the entropy 
from the boost-invariant Bjorken solution (Eq. \ref{eq:sbj})
 (dashed-dotted line). All calculation with $\eta/s=0.2$.}
\label{fig:ent}
\end{figure}
\begin{equation}
S(\tau)= \tau S \int_{-\infty}^\infty d\theta s \cosh(Y-\theta)
\label{eq:stot}
\end{equation}
 increases with time, if shear viscosity is active (Fig. \ref{fig:ent}). 
The increase of the entropy at 
central rapidity $\tau s(\tau,\theta)$
is responsible for an increase of the particle 
multiplicity in the central rapidity region. Estimates of the 
entropy production \cite{Dumitru:2007qr} based on the 
boost-invariant solution (Eq. \ref{eq:bjorkenvis})
\begin{equation}
\frac{d(\tau s)}{d\tau}=\frac{\Pi}{T}
\label{eq:sbj}
\end{equation}
are very close to the 1+1 dimensional
dynamical result,
 it happens if the  shear viscosity is strong enough  to conserve the 
 Bjorken flow in the evolution.

\section{Role of the initial stress tensor}
\label{sec:stressini}

\begin{figure}
\includegraphics[width=.5\textwidth]{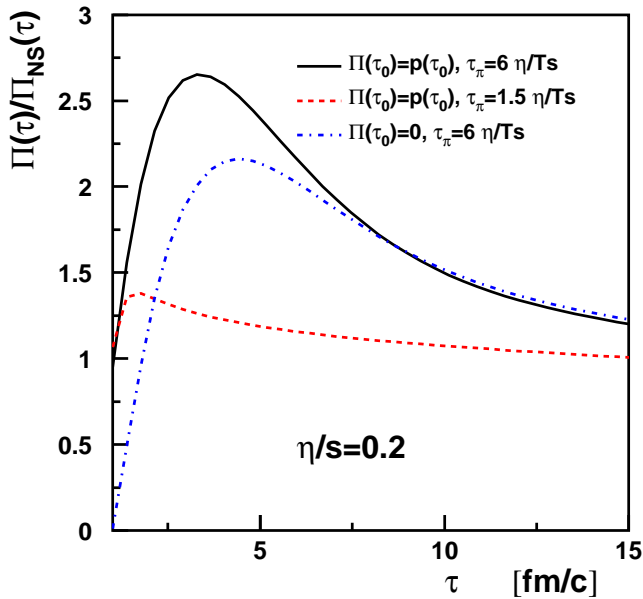}
\caption{(color online) Ratio of the dynamical shear viscosity correction 
$\Pi(\tau)$
 to the Navier-Stokes  value $\Pi_{NS}(\tau)$  at 
$\eta/s = 0.2$  for two initial condition for 
the viscous corrections $\Pi(\tau_0)=p(\tau_0)$ (solid line) and  for 
 $\Pi(\tau_0)=0$ (Eq. \ref{eq:pins}) (dashed-dotted line), and 
for a reduced relaxation time $\tau_\pi=1.5 \eta/T s$ (dashed  line).}
\label{fig:shear}
\end{figure}

Second order viscous hydrodynamics introduces a dynamical equation for the 
stress tensor. A crucial parameter is given by the relaxation time 
$\tau_\pi$. For small relaxation times the viscosity correction stays close 
to the Navier-Stokes value
\begin{equation}
\Pi_{NS}(\tau)=\frac{4 \eta}{3 \tau} \ .
\label{eq:pins}
\end{equation}
Such a behavior of  viscous corrections  has been 
confirmed in numerical simulations of the  transverse expansion in
 viscous hydrodynamics with small relaxation times 
\cite{Romatschke:2007jx,Song:2007fn},
 unlike in Ref. \cite{Bozek:2007di} where a large value of the relaxation
 time was postulated.
We check the effect of the initial value of $\Pi(\tau)$ on the evolution
by comparing two scenarios, $\Pi(\tau_0)=p(\tau_0)$ which corresponds to
an anisotropic momentum (pressure) at the initial time and $\Pi(\tau_0)=0$
 which corresponds to initially locally equilibrated distributions.
With the choice of the the relaxation time $\tau_\pi=6 \eta/T s$ 
the two initial conditions lead to different results. As before, 
for each choice of the initial conditions and parameters we readjust the
 initial energy density distribution (\ref{eq:eini}) to reproduce the final 
pion rapidity distribution. 

\begin{figure}
\includegraphics[width=.5\textwidth]{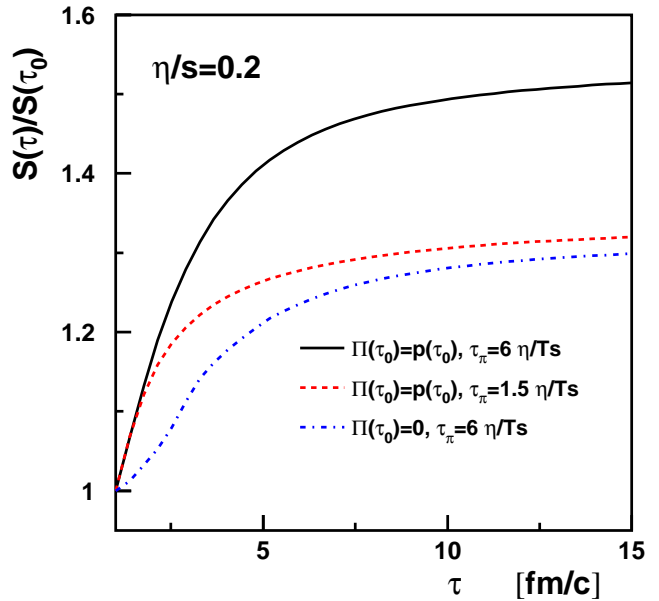}
\caption{(color online) Relative entropy increase in the dissipative evolution
 at 
$\eta/s = 0.2$  for two initial condition for 
the viscous corrections $\Pi(\tau_0)=p(\tau_0)$ (solid line) and for 
 $\Pi(\tau_0)=0$ (Eq. \ref{eq:pins}) (dashed-dotted line), and 
for a reduced relaxation time $\tau_\pi=1.5 \eta/T s$ (dashed line).}
\label{fig:entl}
\end{figure}

In Fig. \ref{fig:shear} is shown the ratio of
 the dynamical value of the viscous correction  $\Pi(\tau)$ to the 
Navier-Stokes value (\ref{eq:pins}). After several fm/c the 
stress tensor, that is set initially to 
zero (dashed-dotted line),
  relaxes to and overshoots the steady flow value (\ref{eq:pins}). 
For an initial stress tensor corresponding to  the anisotropic effective 
pressure (solid line) the 
viscous correction overshoots the Navier-Stokes value  almost immediately. 
Since the dissipative effects are the strongest at the early stages, 
the integrated entropy production is smaller for an evolution
 with initial zero stress tensor 
than for an evolution 
starting with nonzero stress tensor (Fig. \ref{fig:entl}). Even though,
the effect of the change of the initial viscosity correction can be 
counterbalanced by  a suitable change in the initial energy density
 distribution,  the amount of the entropy produced in the dynamics 
depends on the initial value  of the dissipative corrections.
We have also performed a calculation with a smaller relaxation time 
(dashed lines in Fig. \ref{fig:shear} and \ref{fig:entl}). In that 
case the initial value of $\Pi(\tau)$ is less important, as it rapidly 
relaxes to the the steady-flow value $\Pi_{NS}(\tau)$. The integrated 
dissipative effects are smaller.

\section{Expectations for LHC}

\begin{figure}
\includegraphics[width=.5\textwidth]{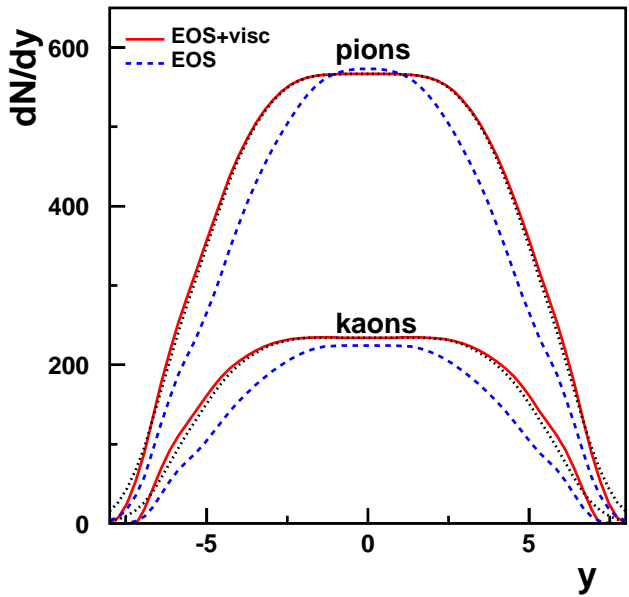}
\caption{(color online) Rapidity distribution of pions (upper curves) and kaons
(lower curves, $\times 3$) for LHC  plateau-like initial conditions
(Eq. \ref{eq:einilhc}), 
  calculated for an evolution with 
shear viscosity coefficient $\eta/s = 0.2$ (solid lines) and for an ideal
 fluid with a realistic equation of state (dashed lines).  The dotted line 
denotes the result of a viscous hydrodynamic evolution,
 but neglecting the viscous corrections to the particle emission at 
freeze-out (Eq. \ref{eq:dndydissi}).}
\label{fig:dndyl}
\end{figure}

\begin{figure}
\includegraphics[width=.5\textwidth]{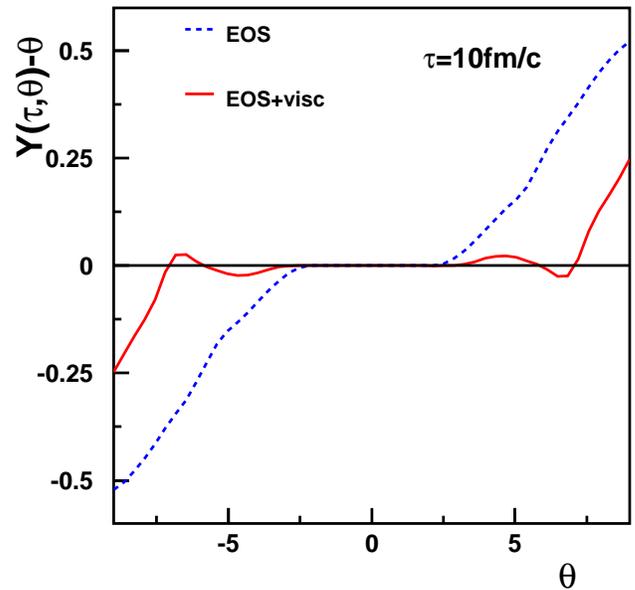}
\caption{(color online)Difference between the  flow rapidity 
of the fluid and the Bjorken value at $\tau=10$fm/c,
 calculated for an evolution with 
shear viscosity coefficient $\eta/s = 0.2$ (solid line) and for an ideal
 fluid with a realistic equation of state (dashed line) for LHC 
 plateau-like initial conditions.}
\label{fig:flowl}
\end{figure}

\begin{figure}
\includegraphics[width=.5\textwidth]{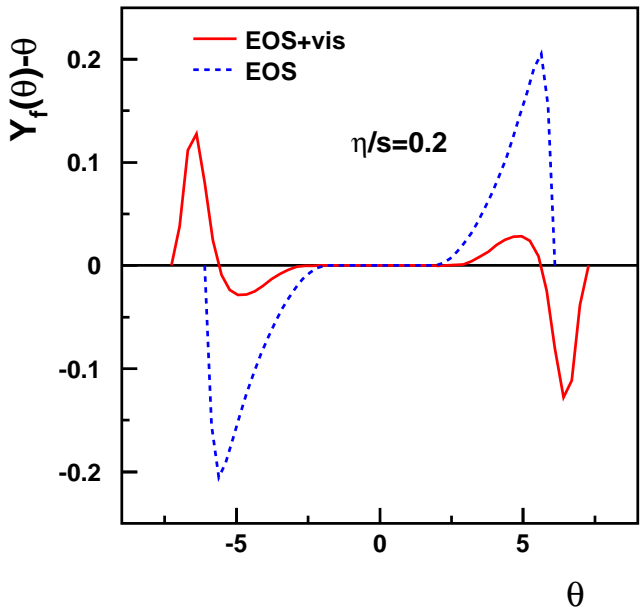}
\caption{(color online) Difference between the  flow rapidity 
of the fluid and the Bjorken value at the freeze-out hypersurface,
 calculated for an evolution with 
shear viscosity coefficient $\eta/s = 0.2$ (solid line) and for an ideal
 fluid with a realistic equation of state (dashed line) for LHC 
 plateau-like initial conditions.}
\label{fig:flowfreezel}
\end{figure}

In this section we present some simple estimates 
of the effects of the shear viscosity on the 
longitudinal expansion at LHC energies. To get a rough estimate
we set arbitrary the multiplicity of pions for central rapidity at 
twice the value observed for central collisions at RHIC.
The initial energy density distribution in space-time rapidity is modified,
it includes a plateau of width $\sigma_p$ \cite{Hirano:2001yi}
\begin{equation}
\epsilon(\tau_0,\theta)=\epsilon_0 
\exp\left(-(\theta-\sigma_p)^2\Theta(|\theta|-
\sigma_p)/2\sigma^2\right) \ .
\label{eq:einilhc}
\end{equation}
The width of the plateau is $\sigma_p=3.3$ and takes all the
 increase of the rapidity range when going from RHIC to LHC energies. The
 parameter $\sigma$ is the same as at RHIC energies, and the energy density
 $\epsilon_0$ is adjusted to reproduce the assumed final pion density 
$\frac{dN}{dy}|_{y=0}$. The meson distributions are shown in Fig. 
\ref{fig:dndyl}. For the viscous evolution the  plateau in the final
 meson distributions survives. 
 For an ideal fluid one gets $\epsilon_0=16.9$GeV/fm$^3$,  
while for a shear viscosity $\eta/s=0.2$ one needs $\epsilon_0=12.4$GeV/fm$^3$.
The difference between the initial energy densities in the viscous
 and ideal fluid evolutions is not as big as for RHIC energies. At LHC energies,
both the ideal and viscous fluid  evolutions have a Bjorken scaling form 
in several units of central rapidity (Fig \ref{fig:flowl}). The lack 
of space-time rapidity gradients in the distributions makes the 
evolution and cooling
 last longer, around $20$fm/c. Even at the freeze-out the flow is Bjorken-like
at central rapidities (Fig. \ref{fig:flowfreezel}). This observation
 justifies  the use at LHC energies of thermal and hydrodynamic models 
assuming boost invariance \cite{Kolb:2000sd,Kisiel:2005hn,Amelin:2006qe}. 
The hydrodynamic evolution can be 
restricted to the  1+2 dimensional boost invariant geometry 
to describe matter 
in the $4$-$5$ central units of rapidity. 
At the freeze-out hypersurface 
 dissipative corrections to the momentum distributions have almost 
disappeared (compare the solid and dotted lines in Fig. \ref{fig:dndyl}).

Since the cooling of the fluid due to the longitudinal expansion is slow 
(like in the Bjorken solution), a realistic modeling of the time 
scales and of the  freeze-out hypersurface must take into account 
the transverse expansion. The speed of the transverse expansion 
would determine the life-time of the system, while at RHIC energies
 longitudinal expansion (in space-time rapidity)  is also important.
We have also performed calculations using Gaussian initial energy 
distributions (\ref{eq:eini}) with rescaled width parameters
 $\sigma$ for the increased  LHC rapidity range.
 For a viscous fluid we find 
a broad region of rapidities where the 
Bjorken scaling flow survives through the 
evolution, $|\theta|<2.5$. For the ideal fluid the scaling is broken, but to
a significantly lesser extend than at RHIC energies.

\section{Conclusions}

The evolution of a fireball of dense and hot matter created in heavy-ion 
collisions can be modelled as a hydrodynamic expansion of a viscous fluid. 
We analyze the effects of the shear viscosity on the longitudinal expansion
of the matter. We solve numerically coupled evolution equations 
for the longitudinal flow, the energy density and the viscous 
corrections in a 1+1 
dimensional geometry, corresponding to a rapid expansion in the beam direction.
As a function of the space-time rapidity the distribution of  matter 
evolves slowly with proper time. The average density drops and the distribution 
gets wider. The last 
phenomenon takes place when the flow of the fluid 
gets stronger than the Bjorken one. 
At the freeze-out temperature the hydrodynamic stage finishes and
particles are emitted thermally according to the Cooper-Frye formula. 
Experimental measurements of the distribution of mesons in rapidity 
\cite{Bearden:2004yx} constrain the allowed distribution of the longitudinal 
velocities of the fluid elements. Correlation between space-time and 
momentum rapidities of the fluid, means that the space-time 
rapidity extension of the fluid must be limited and that its 
longitudinal flow cannot  deviate significantly from the Bjorken flow. 
Shear viscosity counteracts 
the gradients of the velocity field. As
 a consequence it slows down the longitudinal expansion. At the freeze-out 
the energy density distribution in space-time rapidity is narrower and the 
longitudinal flow gets less accelerated than for the ideal fluid hydrodynamics.
Fitting the initial energy density distribution
 to reproduce the final meson distributions, one observes
 a striking effect. With increasing shear 
viscosity coefficient the initial energy 
density of the fireball decreases significantly, 
from $16.9$GeV/fm$^3$ for an ideal fluid to $5.6$GeV/fm$^3$ for  $\eta/s=0.2$.
In the 1+1 dimensional longitudinal geometry, 
this energy density corresponds to an average over the transverse plane.
 Nevertheless, estimates of the maximal energy density reached
 in heavy-ion collisions at RHIC energies must be strongly revised down if 
shear viscosity is effective during the expansion of the fireball. 
This dramatic reduction of the initial density
 should also  be taken into account in hydrodynamic models
dealing with transverse expansion only, both in 1+1 and 1+2 dimensions.

 Depending on 
the balance of the acceleration of the flow from pressure gradients and 
deceleration from viscosity, the flow gets faster or slower than 
the Bjorken one. For some values of the parameters, effects of the
shear viscosity and pressure gradients on the  longitudinal flow of the fluid
 cancel, 
i.e. the flow stays close to the Bjorken flow. This could be an argument 
justifying models, which  combine transverse viscous expansion 
with a Bjorken flow in the beam direction.
When the initial conditions are adjusted to reproduce the final meson
 distributions, we find that the freeze-out hypersurfaces are very similar, 
irrespective of the value of the shear viscosity coefficient, obviously 
the lifetime of the system  is not sensitive to viscous effects either
(Table \ref{table2}). At the freeze-out the viscous corrections 
(from the longitudinal flow) to the thermal
distributions are small, except at large space-time rapidities.
At LHC energies a substantial rapidity plateau, where Bjorken scaling applies, 
is expected to appear. Shear viscosity helps to preserve 
it in a wider rapidity interval 
through
 the evolution.

\section*{Acknowledgments}
The author is grateful to Miko\l aj Chojnacki and Wojtek 
 Florkowski for making available 
the parameterization of the equation of state.
\bibliography{hydr}

\end{document}